\documentclass[prd,showpcs,showkeys,twocolumn]{revtex4}
\usepackage{graphicx}
\usepackage{amsmath, amsfonts, amssymb}
\usepackage{eurosym}
\usepackage{epsfig}
\begin{document}

\title{Formation of Wormholes by Dark Matter in the Galaxy Dragonfly 44}
\author{Sayeedul Islam}
\email{sayeedul.jumath@gmail.com}
\affiliation{Department of Mathematics, Jadavpur University, Kolkata
700032, West Bengal, India}

\author{Ali \"{O}vg\"{u}n}
\email{ali.ovgun@pucv.cl}
\affiliation{Instituto de F\'{\i}sica, Pontificia Universidad Cat\'olica de
Valpara\'{\i}so, Casilla 4950, Valpara\'{\i}so, Chile}
\affiliation{Physics Department, Arts and Sciences Faculty, Eastern Mediterranean University, Famagusta, North Cyprus via Mersin 10, Turkey}
\affiliation{TH Division, Physics Department, CERN, CH-1211 Geneva 23, Switzerland }

\author{Farook Rahaman}
\email{rahaman@associates.iucaa.in}
\affiliation{Department of Mathematics, Jadavpur University, Kolkata
700032, West Bengal, India}

\author{Mustafa Halilsoy}
\email{mustafa.halilsoy@emu.edu.tr}
\affiliation{Physics Department, Arts and Sciences Faculty, Eastern Mediterranean University, Famagusta, North Cyprus via Mersin 10, Turkey}

\date{\today}
\begin{abstract}
Recently, ultra diffuse galaxy (UDG) of Dragonfly 44 in the Coma Cluster was observed and observations of the rotational
speed suggest that its mass is almost same as the mass of the Milky Way. On the other hand, interestingly, the galaxy
emits only 1 \% of the light emitted by the Milky Way. Then, astronomers
reported that Dragonfly 44 may be made almost entirely of dark matter.
 In this study we try to show that the dark matter that constitutes  Dragonfly 44 can form the wormhole or not. Two possible dark matter profiles are used, namely, ultra diffuse galaxy King's model and generalized Navarro-Frenk-White (NFW) dark matter profile. We have shown that King's model dark matter profile does not manage to provide wormhole whereas generalized Navarro-Frenk-White (NFW) dark matter profile is managed to find wormholes.

\end{abstract}
\maketitle

\section{Introduction}

Today, one of the challenging questions in a theoretical physics is
the question of the existence of traversable wormholes \cite{Einstein,moris,moris2,visser1,ao1,rahaman,Richarte:2017iit,Capozziello:2007ec,Delgaty:1994vp,Perry:1991qq,Cataldo:2017yec,Cataldo:2016dxq,Cataldo:2008pm,Jusufi:2017vta,Sakalli:2015taa,Sakalli:2015mka,ao3,metric,outer1,central,einasto}. Other mysteries
in physics is dark matter (DM) \cite{Sahni:2004ai,Feng:2010gw,Albada:1986roa}.
Like black holes there is another miraculous objects of our universe which are wormholes. After the prediction of Einstein-Rosen bridge in 1935 \cite{Einstein}, with a lot of theoretical evidence researchers have proposed the existence of wormholes in space-time which acts like a shortcut path to travel between any two widely separated or infinite region of the universe or between another universes in multi universe model. Structurally it looks like a tunnel (called its throat) with two mouths (most likely spheroidal). Here most interesting thing is exotic matter needed to open its throat (violates null energy conditions \cite{moris,moris2,visser1}. The curiosity about wormholes physics has vigorously been increased since publication of Morris and Thorne's research article where they proposed the prospect of the existence of traversable wormholes, as a solution of Einstein's field equations, that does not contain event horizon and traveler can easily move in both the regions in space-time through its straight stretch throat \cite{moris,moris2}. By the existence of these hypothetical objects one can realize time machines or shortcut among faraway places of space.

Maximum mass of the universe is made up by the most mysterious substance which do not interact with the electromagnetic force, i.e. can't absorb, reflect or emit light are dark  matter whose nature and composition remain  overall question mark \cite{Sahni:2004ai,Feng:2010gw}. Researchers hypothesize the existence of dark matter only from the gravitational effect. Observation of the spiral galaxy rotation curves is an enthralling experimental evidence for the existence of dark matter. The dark matter candidate particles are generally classified into cold, warm and hot categories. Interaction of a scalar field whose energy density is dark energy may be caused of the dark matter particle mass. Over several decades it is openly accepted that almost every galaxy contain a large amount of non luminous  matter forming massive dark matter halos around the galaxy based on different lines of evidence like flat rotation curves of  spiral galaxies \cite{Albada:1986roa} and strong lensing system \cite{Keeton:1997by}.

The model under consideration predicts that the ultra diffuse galaxies (UDGs) space distribution for 90 globular clusters observed around Dragonfly 44 \cite{Dragonfly44}
has the space density:  \\
a.  King's model like 
\citep{king,king2}:
\begin{equation}
\rho(r)\propto \kappa\left(\frac{r}{r_{0}}^{2}+\lambda\right)^{\eta},
\end{equation}
where $\eta$ , $\kappa$, $r_0$  and $\lambda$ are parameters . We assume that the averaged
relative speed of the galaxies in the Coma cluster is approximately
equal to the velocity dispersion in the cluster $v\cong1000$ km/s. This density profile is shown in fig 1.
\\
b. Generalized Navarro-Frenk-White (NFW) profile \cite{Navarro:1995iw}:
\begin{equation}
\rho(r)\propto\frac{1}{r^\gamma\left[1+\left(\frac{r}{r_0}\right)\right]^{3-\gamma}},
\end{equation}
where $\gamma $ and  $r_0$    are parameters .

\begin{figure}[h]
\includegraphics[scale=0.4]{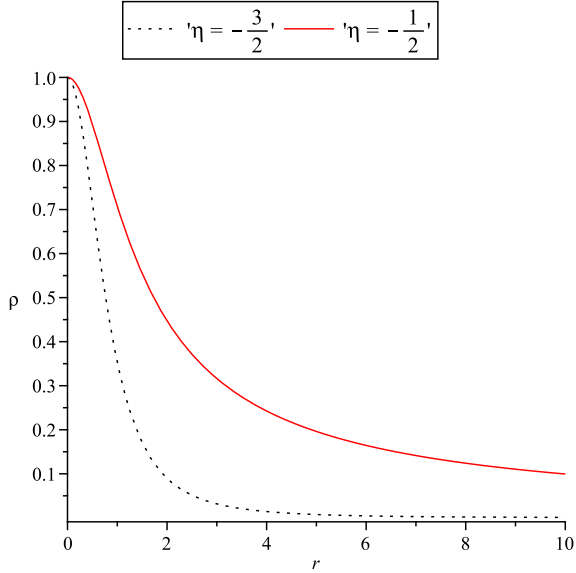}
\caption{Plot $\rho $ versus $r$ (  $\kappa=1$,$\lambda=1$ and $r_0=1$ ).}
\label{Fig. 1}
\end{figure}

This paper is organized as follows. Sec. II we introduces the traversable
wormholes and their basic equations. In Sec. III and IV, we study the possibility
of wormholes in the galaxy of Dragonfly 44 by using the UDGs dark
matter profile \cite{
king,king2,Dragonfly44} and NFWs dark matter profile \cite{Navarro:1995iw},respectively.
Finally, in Sec. V, we present our remarks.

\section{  Wormholes formulation}

The traversable wormhole space-time is given by following \cite{moris,moris2}

\begin{equation}
ds^{2}=-e^{2f(r)}dt^{2}+\left(1-\frac{b(r)}{r}\right)^{-1}dr^{2}+r^{2}(d\theta^{2}+\sin^{2}\theta\,d\phi^{2}).\label{line}
\end{equation}

It is noted that the $b(r)$ and $f(r)$ stands for the spatial shape
function, and for the redshift function, respectively. The range of
the radial coordinate is from +$\infty$ to $b(r_{0})=r_{0}$ where
r$_{0},$ is the minimum value of the r (throat of the wormhole).

The Einstein field equations are given by
\begin{equation}
G_{\nu}^{\mu}=8\pi T_{\nu}^{\mu}
\end{equation}
where $T_{\nu}^{\mu}$ and $G_{\nu}^{\mu}$ are the stress-energy
tensors and the Einstein tensor, respectively. Then one can calculate
the non-zero Einstein tensors:
\begin{equation}
G_{t}^{t}=\frac{b^{\prime}}{r^{2}},\label{gtt}
\end{equation}
\begin{equation}
G_{r}^{r}=\frac{-b}{r^{3}}+2\left(1-\frac{b}{r}\right)\frac{f^{\prime}}{r},
\end{equation}
\begin{equation}
G_{\theta}^{\theta}=\left(1-\frac{b}{r}\right)\left[f^{\prime\prime}+{f^{\prime}}^{2}+\frac{f^{\prime}}{r}-\left(f^{\prime}+\frac{1}{r}\right)\left\{ \frac{b^{\prime}r-b}{2r(r-b)}\right\} \right],
\end{equation}
\begin{equation}G_{\phi}^{\phi}=G_{\theta}^{\theta} \end{equation}
where a prime is $\frac{d}{dr}.$

DM is generally defined in the form of general anisotropic energy-momentum
tensor
\begin{equation}
T_{\nu}^{\mu}=(\rho+p_{r})u^{\mu}u_{\nu}+p_{r}g_{\nu}^{\mu}+(p_{t}-p_{r})\eta^{\mu}\eta_{\nu},
\end{equation}
where $u^{\mu}u_{\mu}=-\frac{1}{2}\eta^{\mu}\eta_{\mu}=-1$. Note
that $p_{t}$ stands for the transverse pressure, $p_{r}$ is the
radial pressure and $\rho$ is the energy density. A possible set
of $u^{\mu}$ and $\eta^{\mu}$ are given by $u^{\mu}=(e^{2f(r)},0,0,0)$
and $\eta^{\mu}=(0,0,\frac{1}{r},\frac{1}{r\sin\theta})$. Then the
stress-energy tensors $T_{\nu}^{\mu}$ are calculated as follows
\begin{equation}
T_{t}^{t}=-\rho,\label{Tr}
\end{equation}
\begin{equation}
T_{r}^{r}=p_{r},
\end{equation}

\begin{equation}
T_{\theta}^{\theta}=T_{\phi}^{\phi}=p_{t}.\label{Ttheta}
\end{equation}
\section{Wormholes with the UDG King's model  DM }

  One can find the  tangential  velocity    from the flat rotation curve for the
circular stable geodesic motion in the equatorial plane as \cite{einasto}
\begin{equation}
v^{\phi}=\sqrt{rf^{\prime}}\label{v1},
\end{equation}
which is responsible to fit the flat rotational curve for the DM.
Rahaman et. al observe the rotational curve profile in   the DM region
as follows \cite{central,outer1}
\begin{equation}
v^{\phi}=\alpha r\exp(-k_{1}r)+\beta\lbrack1-\exp(-k_{2}r)]\label{v2}
\end{equation}
where $\alpha,$ $\beta,$ $k_{1},$ and $k_{2}$ are constant positive
parameters.

Using the Eqns.(\ref{v1}) and (\ref{v2}), the redshift function
is obtained as follows
\begin{eqnarray}
f(r) &=&-\frac{\alpha ^{2}r}{2k_{1}e^{(2k_{1}r)}}-\frac{\alpha ^{2}}{%
4k_{1}^{2}e^{(2k_{1}r)}}-\frac{2\alpha \beta }{k_{1}e^{(k_{1}r)}} \notag
\\
&&+\frac{%
2\alpha \beta e^{(-k_{1}r-k_{2}r)}}{k_{1}+k_{2}} +\beta ^{2}\ln (r)+2\beta ^{2}E_{i}(1,k_{2}r) \notag
\\
&&-\beta ^{2}E_{i}(1,2k_{2}r)+D.
\label{fr}
\end{eqnarray}%
where E$_{i}$ and $D$ are the exponential integral and integration
constant, respectively. At large scales, it becomes $e^{2f(r)}=B_{0}r^{(4v^{\phi})}$.\\

Now we discuss two cases for different values of the parameter $\eta$ by using UDGs king's dark matter profile for the stability of wormholes in the galaxy Dragonfly 44.
\subsection{The case I}

For the UDGs King's DM density profile \cite{
king,king2,Dragonfly44}
\begin{equation}
\rho(r)= \kappa\left[\left(\frac{r}{r_{0}}\right)^{2}+\lambda\right]^{\eta},
\end{equation}
we assume in this case  $\kappa=1$, $\lambda=1$ and $\eta=-3/2$.
After one uses the Eq.s (\ref{v1}) and (\ref{v2}) and the UDG's
dark matter density profile under the Einstein field equations, the
shape function is calculated ($8\pi=1$) as
\begin{equation}
b(r)=\frac{rr_{0}^{2}}{8(r^{2}+r_{0})}+\frac{r_{o}^{3/2}}{8} \tan^{-1}\left(\frac{r}{\sqrt{r_{0}}}\right)-\frac{r_{0}^{3}r}{4(r^{2}+r_{0})^{2}}+C
\end{equation}
Note that $C$ is the integration constant and it is chosen as
\begin{equation}
C=r_{0}-\frac{r_{0}^{2}}{8(1+r_{0})}+\frac{r_{o}^{3/2}}{8}\tan^{-1}(\sqrt{r_{0}})+\frac{r_{0}^{2}}{4(1+r_{0})^{2}}\label{constant}
\end{equation}
to satisfy the condition of $b(r_{0})=r_{0}$, then it is checked
the flare-out condition ($b^{\prime}<1$), where $r_{0}$ is the radius of throat.
\begin{equation}
b^{\prime}=\rho r^{2}=\frac{r_{0}^{3}r^{4}}{(r^{2}+r_{0})^{3}}\label{flr}
\end{equation}
which is satisfied (see fig.(2)).

\begin{figure}[h]
\includegraphics[scale=0.4]{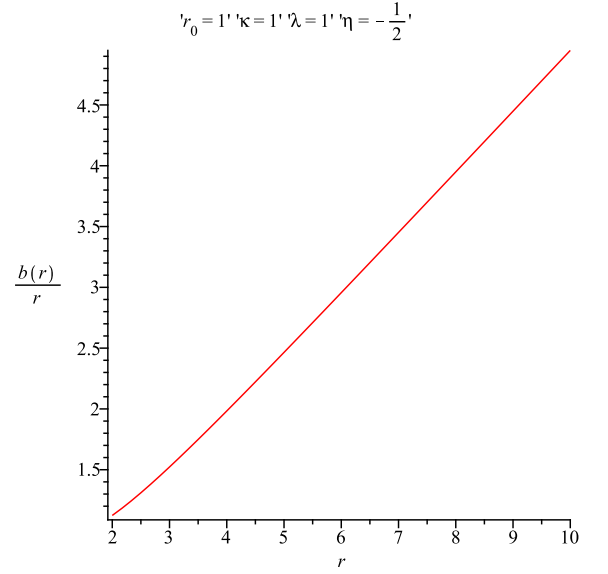}
\caption{The figure is shown for $b(r)/r$  versus $r$.}
\label{Fig. 2}
\end{figure}

\begin{figure}[h]
\includegraphics[scale=0.4]{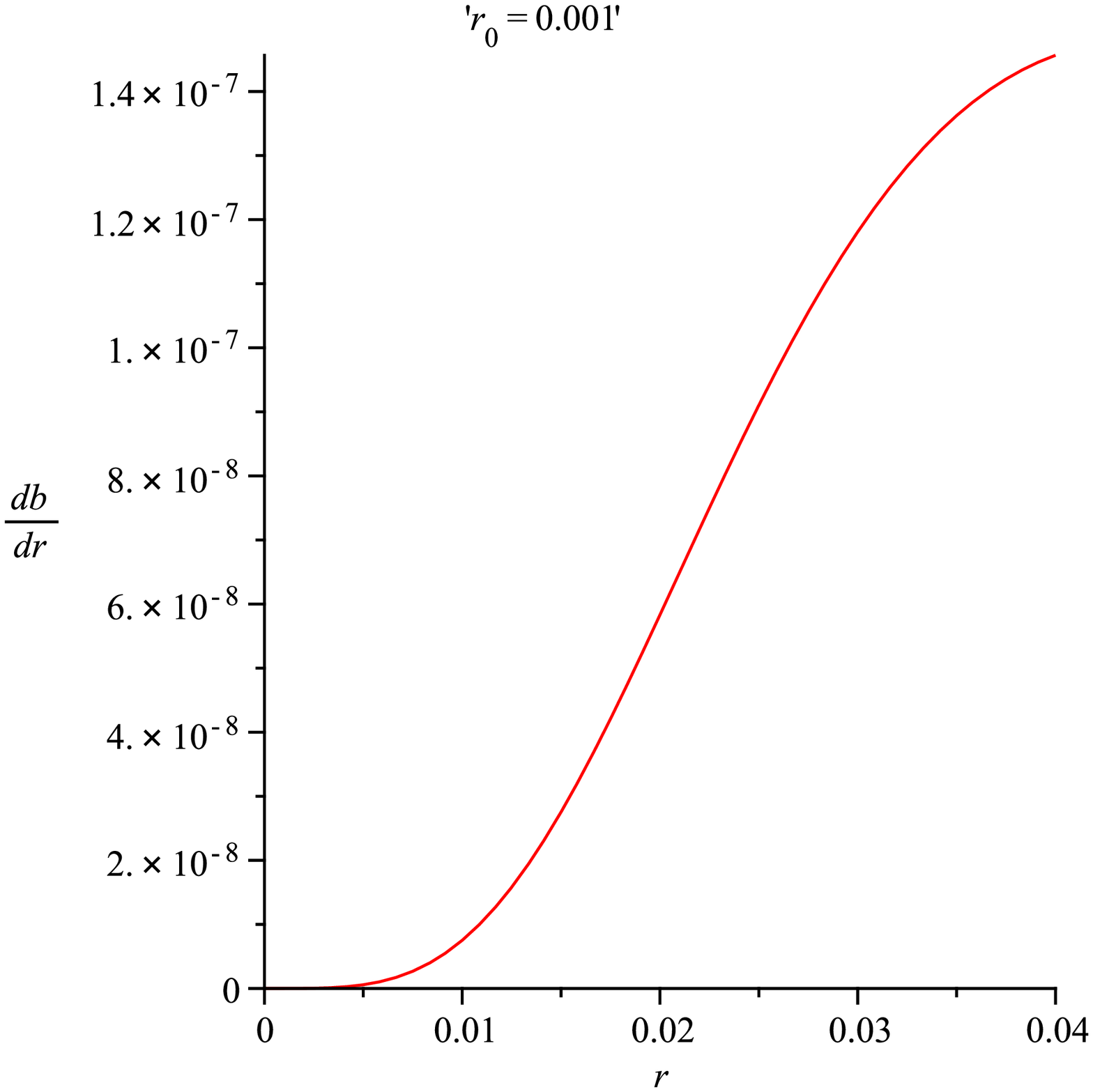}
\caption{The figure is shown for $b^\prime(r)$  versus $r$.}
\label{Fig. 3}
\end{figure}

 Furthermore from the Eq.(\ref{Tr}) the second
derivative of $b$ with respect to $r$ is calculated as

\begin{equation}
b^{\prime\prime}=-\frac{2r_{0}^{3}r(2r^{2}-r_{0})}{(r^{2}+r_{0})^{4}}.
\end{equation}

The radial of pressure is showed as follows by substituting $f(r)$
and $b(r)$ into the solution (Eqns. 5- 12):

\begin{equation}
p_{r}=\frac{-b}{r^{3}}+2\left(1-\frac{b}{r}\right)\frac{f^{\prime}}{r},
\end{equation}

\begin{eqnarray}
f^{\prime}={\frac{{\alpha}}{{2}r}{ \left( {{\rm e}^{{k_{1}}\,r}} \right) ^{2}}
}-2\,{\alpha}\,{\beta}\,{{\rm e}^{-r \left( {k_{1}}+{k_{2}}
 \right) }}+2\,{\frac {{\alpha}\,{\beta}}{{{\rm e}^{{k_{1}}\,r}}}} \notag
&&\\+{
\frac {{{\beta}}^{2}{{\rm e}^{-2\,{k_{2}}\,r}}}{r}} -2\,{\frac {{{
\beta}}^{2}{{\rm e}^{-{k_{2}}\,r}}}{r}}+{\frac {{{\beta}}^{2}}{r}}.
\end{eqnarray}
It is showed in the figure (4)  that the null energy condition ($\rho+p_{r}<0$)
is violated so that one of the essential condition for a wormhole is satisfied. However, the most important criterion  for wormhole, namely, $\frac{b(r)}{r}  <1$ is violated (see fig.2). So in this case wormhole does not exists.
\begin{figure}[h]
\includegraphics[scale=0.4]{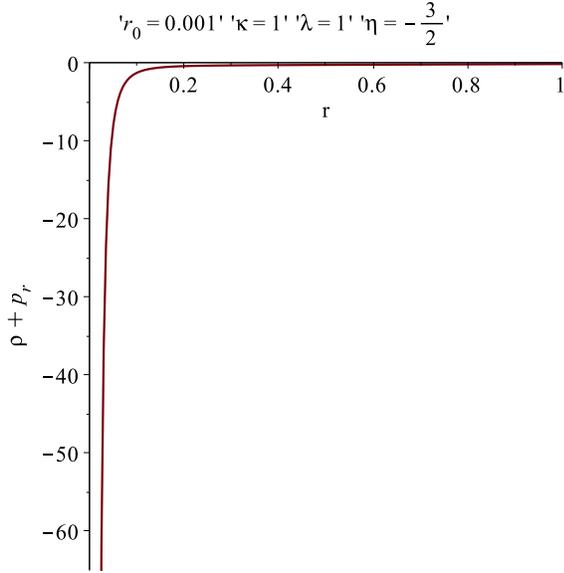}
\caption{Null energy condition for the case I.}
\label{Fig. 4}
\end{figure}

\subsection{The case II}

In this section we follow the same method using in case I and  the shape function is calculated .

\begin{figure}[h]
\includegraphics[scale=0.4]{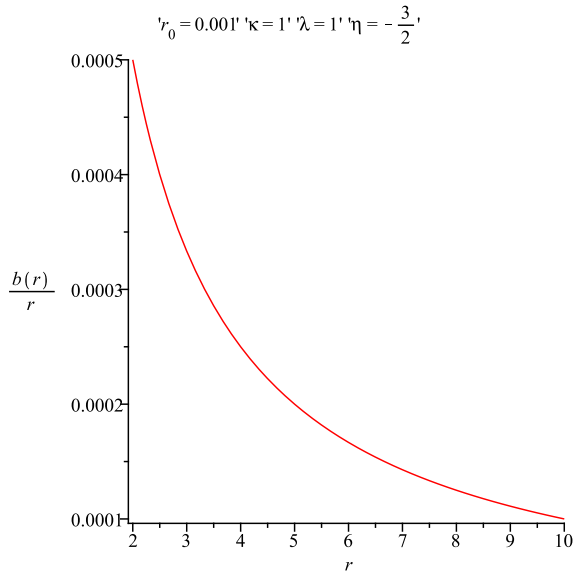}
\caption{The figure is shown for $b(r)/r$  versus $r$.}
\label{Fig. 5}
\end{figure}

\begin{figure}[h]
\includegraphics[scale=0.4]{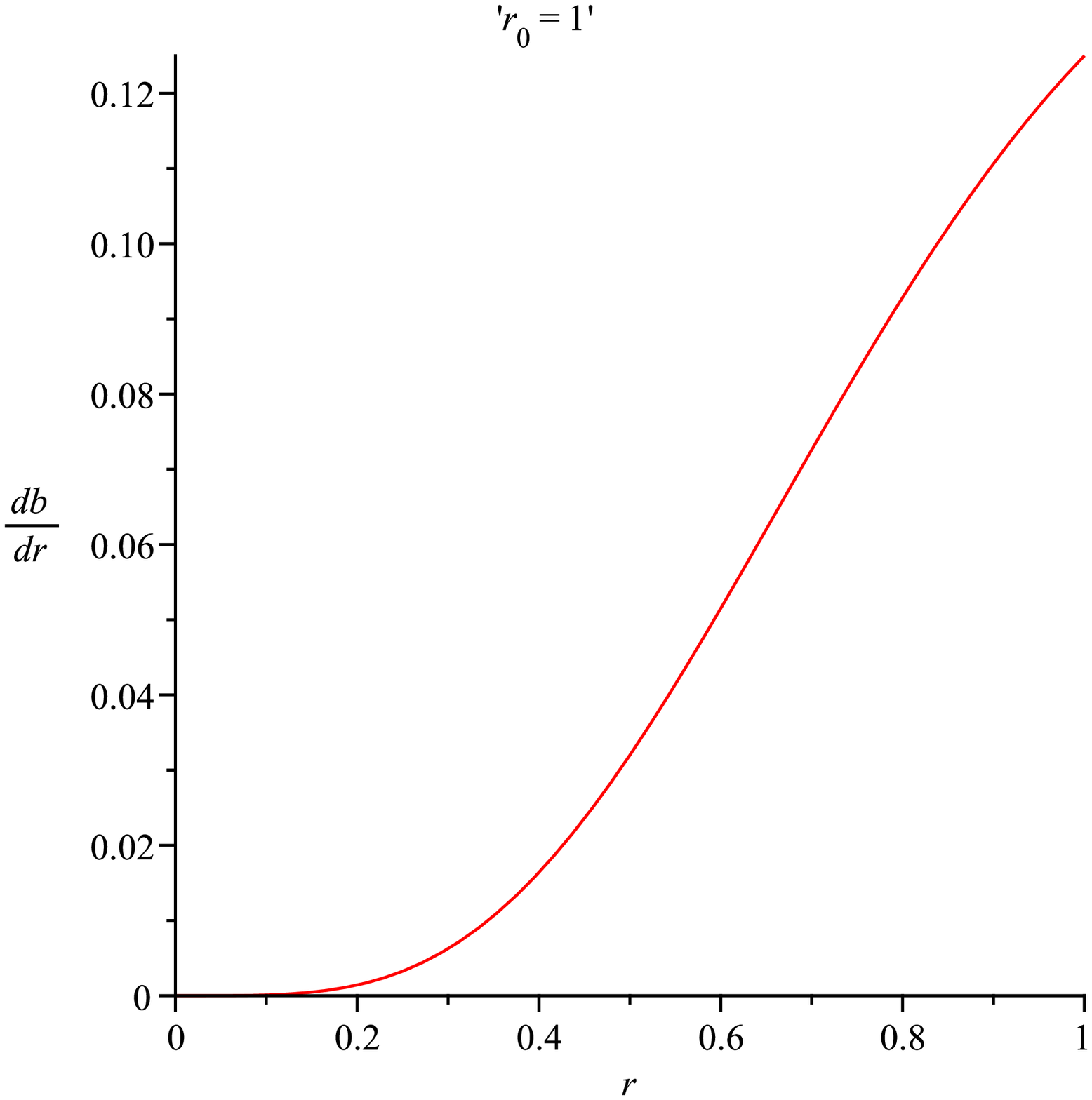}
\caption{The figure is shown for $b^\prime(r)$  versus $r$.}
\label{Fig. 6}
\end{figure}
\begin{figure}[h]
\includegraphics[scale=0.4]{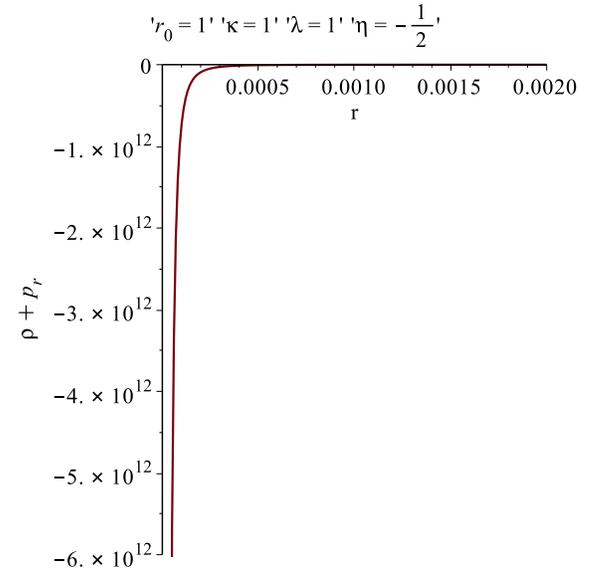}
\caption{Null energy condition for the case II.}
\label{Fig. 7}
\end{figure}

It is checked the flare-out condition ($b^{\prime}<1$) is satisfied  and plotted in Fig. (6).

It is also  showed in the Fig. (7) that the null energy condition ($\rho+p_{r}<0$)
is violated as needed to hold a wormhole. However, as in case I, $\frac{b(r)}{r}  <1$ is violated(see fig.5), so in this case wormhole does not exists. Thus King's model dark matter profile that constitutes  Dragonfly 44 does not manage to provide wormhole. Now we will perform a study whether generalized NFW dark matter profile that constitutes  Dragonfly 44 does   manage to provide wormhole or not.

\section{Wormholes with the NFWs DM }
In this section we describe the dark matter halo distribution using the following generalized Navarro-Frenk-White (NFW) profile for the existence of wormholes in the galaxy Dragonfly 44.
\begin{equation}
\rho(r)= \frac{1}{r^\gamma\left[1+\left(\frac{r}{r_0}\right)\right]^{3-\gamma}}
\end{equation}
In this expression, $\gamma$ is the inner slope of the profile ($\gamma=1$ corresponds to the case of a standard NFW profile), and $r_0$ is the scale radius and taking variation constant as unity.\\

Here we have discussed three different cases for the several values of the parameter $\gamma$

\subsection{case I}
In this case we assume  $r_0=10,\rho_0=0.05$ and $\gamma = 3$.\\
After using NFW dark matter density profile under the Einstein field equations, the shape function $b(r)$ is calculated $(8\pi=1)$ as
\begin{equation}
b(r)=8\pi\rho_0\left[ln(r)+C\right]
\end{equation}
Note that $C$ is the integration constant and it is choosen as
\begin{equation}
C=\frac{r_0}{8\pi\rho_0}-lnr_0\label{constant}
\end{equation}
\begin{figure}[h]
\includegraphics[scale=0.4]{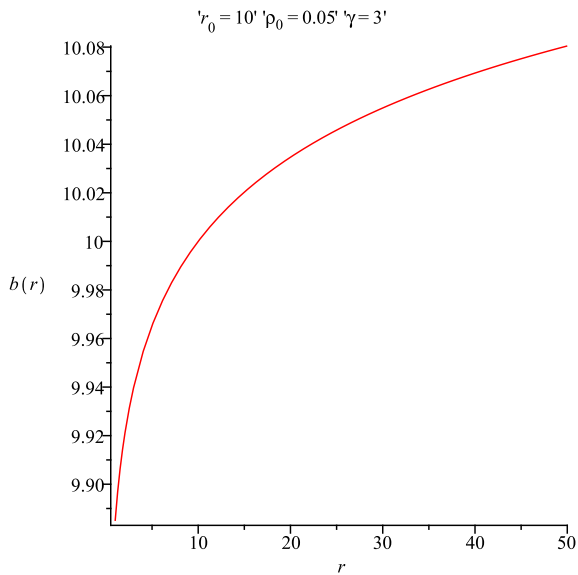}
\caption{ The figure is shown for $b(r)$ versus $r$.}
\label{Fig. 8}
\end{figure}
\begin{figure}[h]
\includegraphics[scale=0.4]{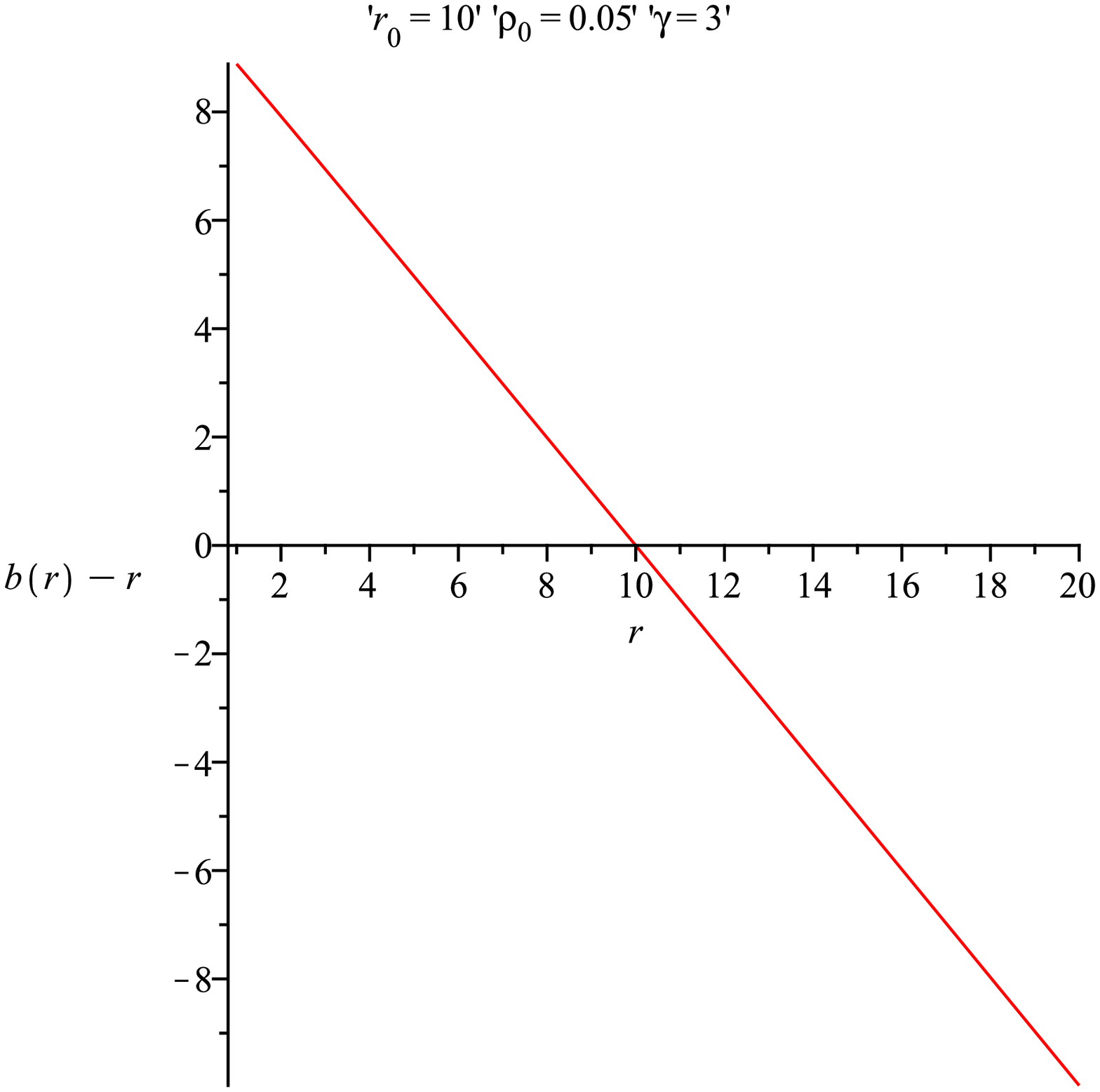}
\caption{ The figure is shown for $b(r)-r$ versus $r$.}
\label{Fig. 9}
\end{figure}
\begin{figure}[h]
\includegraphics[scale=0.4]{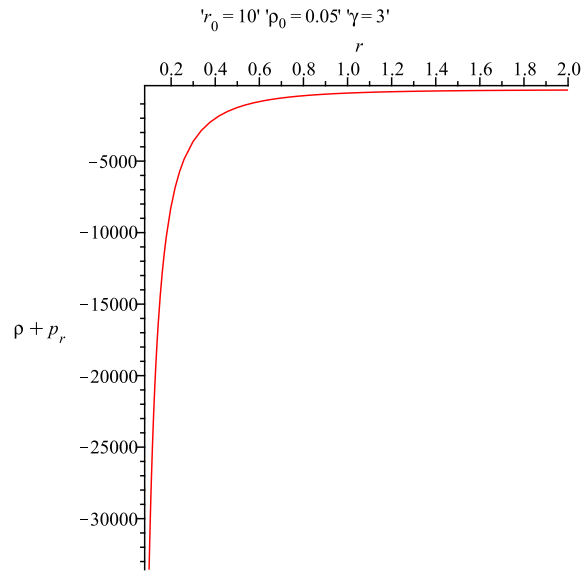}
\caption{ Null energy condition for the case I.}
\label{Fig. 10}
\end{figure}

Here fig.(8) represents the shape function $b(r)$ and we see that the null energy condition ($\rho+p_{r}<0$) is violated ( see fig.10).   Also we have  checked the most important flare-out condition ($b(r)-r < 0$, after the throat radius i.e. $b^{\prime}<1$) which is satisfied and  is plotted in the fig.(9).\\
Thus the generalized NFW dark matter profile that constitutes Dragonfly 44 can manage to provide wormhole.
\pagebreak
\subsection{case II }
In this case
we choose $r_0=10,\rho_0=0.05$ and $\gamma = 4$\\
After using NFW dark matter density profile under the Einstein field equations, the shape function $b(r)$ is calculated $(8\pi=1)$ as
\begin{equation}
b(r)=8\pi\rho_0\left[\frac{-1}{r}+\frac{ln(r)}{r_0}+C\right]
\end{equation}
Note that $C$ is the integration constant and it is choosen as
\begin{equation}
C=\frac{r_0}{8\pi\rho_0}+\frac{1}{r_0}-\frac{1}{r_0}lnr_0\label{constant1}
\end{equation}
\begin{figure}[h]
\includegraphics[scale=0.4]{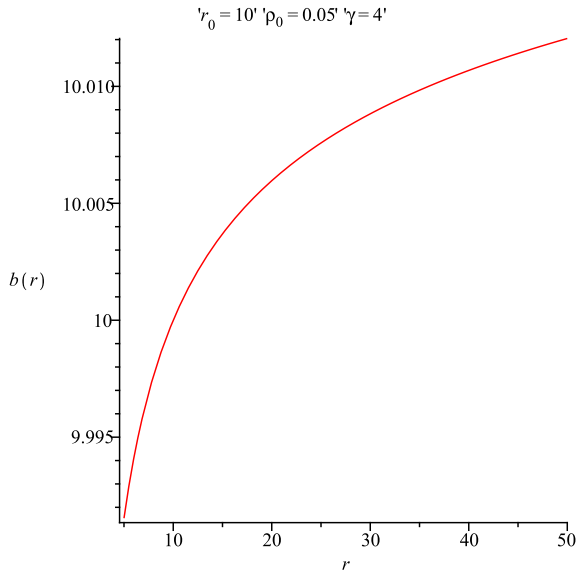}
\caption{ The figure is shown for $b(r)$ versus $r$.}
\label{Fig. 11}
\end{figure}
\begin{figure}[h]
\includegraphics[scale=0.4]{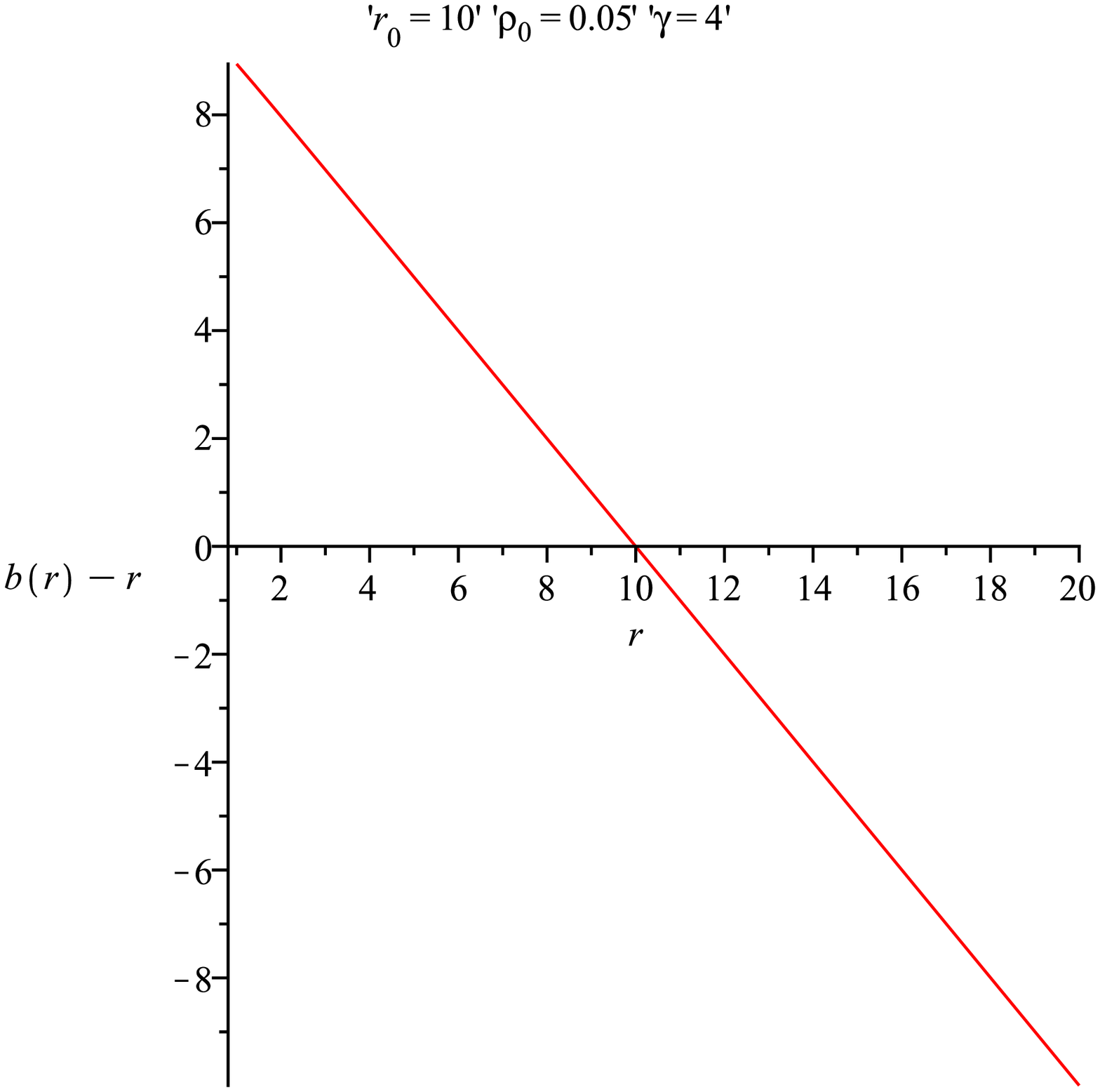}
\caption{ The figure is shown for $b(r)-r$ versus $r$.}
\label{Fig. 12}
\end{figure}
\begin{figure}[h]
\includegraphics[scale=0.4]{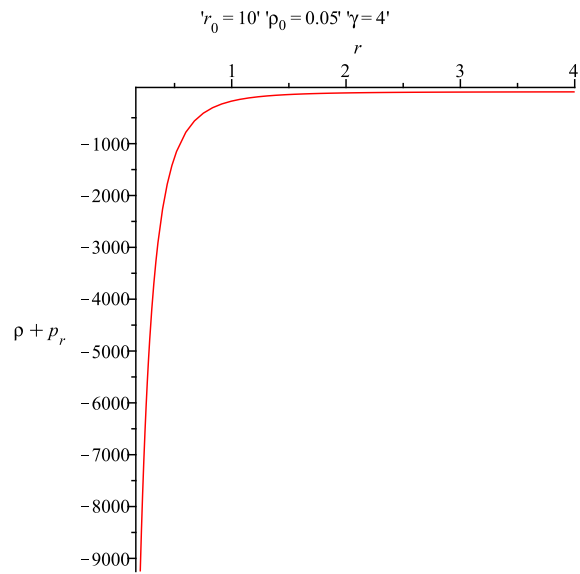}
\caption{ Null energy condition for the case II.}
\label{Fig. 13}
\end{figure}
In this section we follow the same method used in case I and the shape function $b(r)$ is   plotted in the fig. (11). To form a wormhole, the essential criteria regarding  flare-out condition ($b(r)-r < 0$, after the throat radius i.e. $b^{\prime}<1$)  and  violation of null energy condition  ($\rho+p_{r}<0$) are    satisfied  ( see  fig.12 and fig.13 ).\\

\subsection{case III }
In this case we assume
  $r_0=10,\rho_0=0.05$ and $\gamma = 5$\\
After using NFW dark matter density profile under the Einstein field equations, the shape function $b(r)$ is calculated $(8\pi=1)$ as
\begin{equation}
b(r)=8\pi\rho_0\left[\frac{-2}{rr_0}+\frac{ln(r)}{r_0^{2}}-\frac{1}{2r^{2}}+C\right]
\end{equation}
Note that $C$ is the integration constant and it is choosen as
\begin{equation}
C=\frac{r_0}{8\pi\rho_0}+\frac{2}{r_0^{2}}-\frac{1}{r_0^2}lnr_0+\frac{1}{2r_0^{2}}\label{constant}
\end{equation}
\begin{figure}[h]
\includegraphics[scale=0.4]{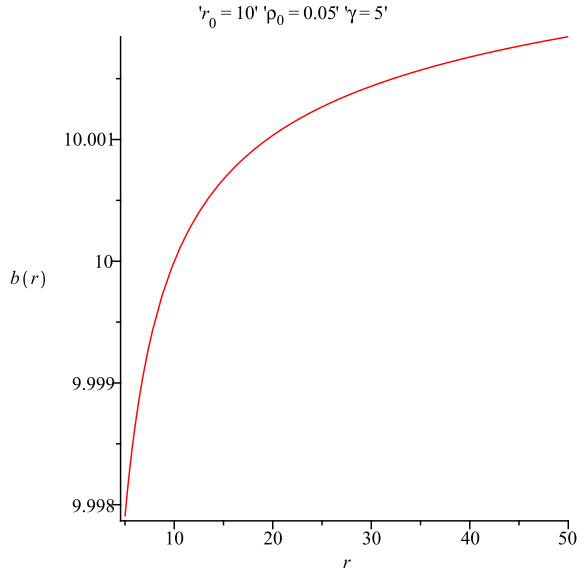}
\caption{ The figure is shown for $b(r)$  versus $r$.}
\label{Fig. 14}
\end{figure}

\begin{figure}[h]
\includegraphics[scale=0.4]{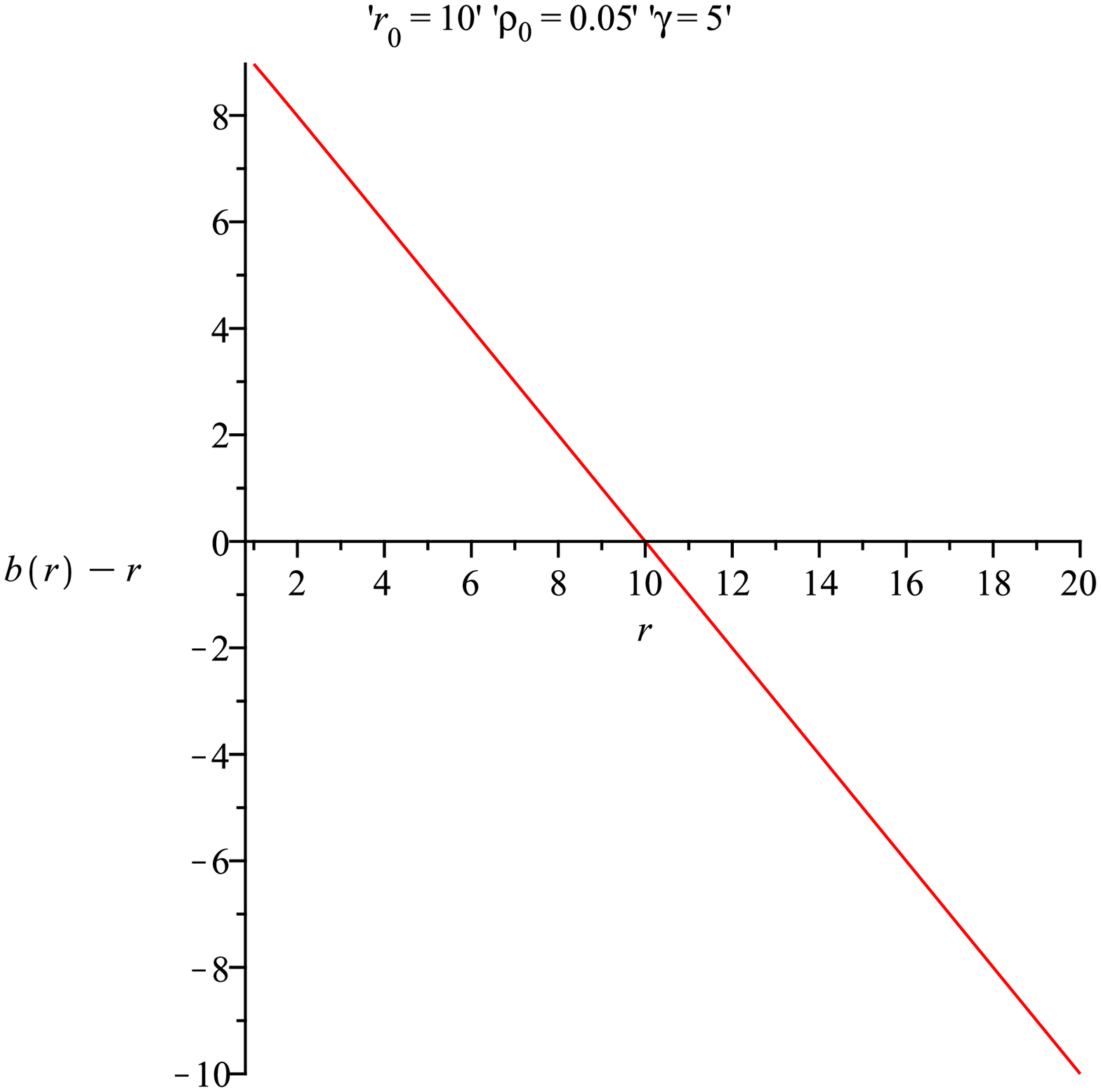}
\caption{ The figure is shown for $b(r)-r$  versus $r$.}
\label{Fig. 15}
\end{figure} 

\begin{figure}[h]
\includegraphics[scale=0.4]{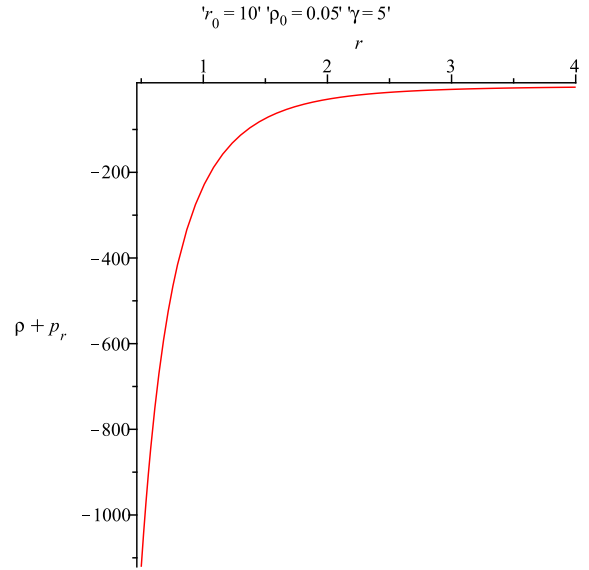}
\caption{Null energy condition for the case III.}
\label{Fig.16}
\end{figure}

Here we use the same NFW dark matter density profile like above cases and similarly the shape function $b(r)$ is calculated and plotted in the fig. (14). \\
In the fig. (15) and fig. (16),  it is clearly shown that the flare-out condition (($b(r)-r < 0$, after the throat radius i.e. $b^{\prime}<1$)) is satisfied and null energy condition  ($\rho+p_{r}<0$) is violated to hold  a   wormhole open.

So we can claim that the generalized NFW dark matter profile that constitutes Dragonfly 44  provides wormhole.
\pagebreak
\section{Conclusion}

The presence of stable traversable wormholes is a noteworthy issue in theoretical physics. There is doubtlessly that wormholes is the most interesting objects in universe. This work is persuaded primarily by Ref.s \cite{outer1,central,einasto}.  In this paper, we use the UDG systems in the Coma Cluster which is known as Dragonfly 44 that astronomers
reported that Dragonfly 44 may be made almost entirely of dark matter. Moreover, there is another possibility: in this study we show that this dark matter can form the wormhole, and it affects the observations. All the normal matter might be passed through this wormhole. For this purpose we firstly use the UDG profile and try to construct wormhole solution, then we repeat our calculations for the Navarro-Frenk-White (NFW) profile. We have shown that  only Navarro-Frenk-White (NFW) profile provide wormhole solutions.   Thus we able to find the solutions of the wormhole in the Dragonfly 44 galaxy so that the dark matter`s halos around the Dragonfly 44 galaxy is suitable to harbor wormholes.

\begin{acknowledgments}
This work was supported by the Chilean FONDECYT Grant No. 3170035 (A\"{O}). A\"{O} is grateful to the CERN theory (CERN-TH) division for hospitality where part of this work was done. FR would like to thank the authorities of the Inter-University Centre
for Astronomy and Astrophysics, Pune, India for providing research facilities.   FR and SI are also grateful to DST-SERB and DST-INSPIRE,  Govt. of India ,  for financial support respectively.
\end{acknowledgments}

\end{document}